Accepted in APPLIED OPTICS

# Evolutions of ring Airy vortex beam and ring Pearcey vortex beam in turbulent atmosphere and a comparative analysis of their channel efficiencies and OAM spectra

SHAKTI SINGH,[1] SANJAY KUMAR MISHRA,[2] AKHILESH KUMAR MISHRA[1,3,*]

[1]Department of Physics, Indian Institute of Technology Roorkee, Roorkee-247667, India
[2]Adaptive Optics Group, Instruments Research and Development Establishment, Dehradun-248008, India
[3]Centre for Photonics and Quantum Communication Technology, Indian Institute of Technology Roorkee, Roorkee- 247667, Uttarakhand, India
*Corresponding author- akhilesh.mishra@ph.iitr.ac.in

**Abstract:** An optical vortex beam propagating through turbulent atmosphere encounters distortions in the wavefront that results in modal scattering. Abruptly autofocussing (AAF) beams with orbital angular momentum have gained significant attention due to their non-diffracting and self-healing nature. These warrants understanding of the behaviour of these beams through turbulent atmosphere absolutely necessary. With this intuition, in the present work we investigate the behaviour of two AAF beams namely ring Airy vortex beam (RAVB) and ring Pearcey vortex beam (RPVB) through the turbulent atmosphere in two cases- multiplexed and non-multiplexed. We propagate multiplexed as well as non-multiplexed RAVB and RPVB in different levels of turbulent atmosphere. In non-multiplexed case, channel efficiency declines for both the beams with increase in modes numbers. In multiplexed case, increasing the gap between the mode sets results in decrease in channel efficiency. We also report that in weak atmospheric turbulence RAVB outperform RPVB in terms of channel efficiency. We use optical transformation sorting (log-polar) method to demultiplex the optical beams at the output. Furthermore, we investigate and compare the OAM spectra of both beams in different levels of atmospheric turbulence and at different propagation distances. The comparison reveals that the spectra of RPVB are more dispersive as compared to that of RAVB.

## 1. Introduction:

Free-space optical communication (FSO) has emerged as an underlying way of transmission and a thriving area of research, due to its remarkable data carrying capacity compared to the conventional radio wave communication [1]. FSO differs from fiber-based communication system by operating without physical cables and utilizing the atmosphere as transmission medium. FSO becomes valuable in situations where fiber-based communication is either not available or disrupted by the natural disasters. Although we have radio frequency (RF) for the transmission of information through turbulence media, but FSO have advantages over RF such as higher bandwidth, higher information carrying capacity, smaller beam divergence and improved security [2]. Modern day data demands rely on the simultaneous transmission of data across multiple channels. This requirement is fulfilled through the utilization of orthogonality of optical beam parameters. In FSO, the data is controlled and multiplexed using different light parameters, including polarization and frequency etc [3]. This control allows for the simultaneous transmission of multiple channels.

Spatial multiplexing opens up another avenue, where spatially orthogonal beams provide new degree of freedom for multiplexing and therefore increase the capacity of the optical channels further. Optical beams with orbital angular momentum (OAM) have been extensively exploited for this type of multiplexing. An OAM beam carries azimuthally varying wavefront described by exp $(il\varphi)$, where $l$ is the topological charge of the optical beam [4-6]. Utilizing the multiplexing capabilities of OAM beam, ultra-high spectrally efficient and petabit scale free space data transmission has been achieved [7,8]. However, there are some challenges in the transmission of these beams that may degrade the performance of OAM based communication. Among them divergence of OAM beams is of primary concern. Since higher order OAM modes diverge faster, usually lower order modes are employed for large distance communication. FSO based on OAM faces random refractive variations which distort the helical wavefront of the OAM beams. The distorted wavefront leads to OAM spectrum dispersion [9-10]. To overcome such challenges, an optical beam with self-healing properties and non-diffracting characteristics is needed for FSO.

Abruptly autofocusing (AAF) beams are a class of such beams which preserves aforementioned properties [11]. Ring Airy vortex beams (RAVB) and ring Pearcey vortex beam (RPVB) are among AAF beams that suit the FSO application and hold implications in various other fields as well [12–15]. Recently, propagation dynamics of a variety of structured light beams have been studied through atmospheric turbulence of different strengths [16–18]. Importantly, in case of Bessel-Gauss and Laguerre Gauss (LG) beams through atmospheric turbulence, it has been noticed that Bessel-Gauss beam outperforms LG beam in term of channel efficiency and bit error rate [19]. In turbulent atmosphere, quasi RAVB beam performs better in comparison to conventional OAM and Bessel beams [20]. Recently, the OAM spectra of LG beam and Bessel beam have been explored and it has been reported that LG beam performs better as compared to the Bessel beam in terms of OAM spectra under turbulent environment [21].

In the present work, we numerically study the dynamics of RAVB and RPVB in atmospheric turbulence. To mimic the turbulent atmosphere numerically, we have used multiple random phase screen methods which rely on Von Karman spectral density function. The major contribution of this paper is the comparison between different performance parameters vital to FSO of the two beams. We have explored several aspects of numerical simulations such as random phase screen generation using subharmonic correction to the spatial frequency, optical transformation sorting (log-polar method) and multiplexing of the optical beams. The paper is organized as follows- in section 2, we have discussed the numerical model of the study. Section 3 introduces the mathematical expressions of the RAVB and RPVB and the simulation parameters used in the numerical experiment. In section 4, we have discussed the optical transformation sorting method. The evolutions of transverse intensity profiles and phases of the multiplexed optical beams are detailed in section 5. In section 6, we present the channel efficiency of nonmultiplexed and multiplexed optical beams. Section 7 discusses the evolutions of OAM spectra of both RAVB and RPVB. Finally, we conclude the study in section 8.

## 2. Numerical modelling of turbulent atmosphere:

Optical beam propagation through the atmosphere encounters three atmospheric processes such as absorption, scattering and refractive index fluctuations [22]. Among these refractive index fluctuation gives rise to intensity fluctuation in the optical beam characterized by a physical quantity named as scintillation index (SI) [23]. The fluctuations are stochastic in nature and therefore the propagation of optical beam through turbulent atmosphere is modelled by stochastic Helmholtz equation. The mathematical expression of stochastic Helmholtz equation is given by,

$$\nabla^2 U + n^2(r)k^2 U = 0, \tag{1}$$

where $k$ is the wavevector constant and refractive index $n$ is a random function of $r$. The random nature of refractive index is characterized by power spectral density function (PSD) [23]. This PSD describes the statistical distribution and size of the turbulent eddies. Turbulent eddies are randomly distributed regions of high and low refractive index. The refractive index variations induced by atmospheric turbulence are minute, therefore optical beam passing through turbulent eddies predominantly undergoes phase changes, with negligible amplitude changes. This random phase variation leads to the appearance of speckle in the optical beam. The extended atmospheric medium is characterized numerically as an ensemble of multiple random phase screens arranged along the propagation direction at equal intervals [24]. The random phase screen mimics turbulent atmosphere by perturbing the phase of the optical beam… Although different types of PSD have been defined, we have used modified Von Karman model to generate the random phase screen. Its mathematical expression is given by [16]

$$\phi_n(\kappa) = 0.033 C_n^2 (\kappa^2 + \kappa_0^2)^{-11/6} \exp\left(\frac{-\kappa^2}{\kappa_m^2}\right), \tag{2}$$

where $C_n^2$ represents the refractive index structure constant that represents the strength of atmospheric turbulence. Its typical value varies from $C_n^2 = 10^{-17} m^{-2/3}$ to $C_n^2 = 10^{-12} m^{-2/3}$ representing transition from weak to strong turbulence. In the above equation, $\kappa_0 = \frac{2\pi}{L_0}$ and $\kappa_m = \frac{5.92}{l_0}$, where $L_0$ represents the outer scale and $l_0$ represents the inner scale of the atmospheric turbulence. When the turbulent eddies lie between $L_0$ and $l_0$, the turbulent atmosphere is considered as homogeneous and isotropic. Close to the Earth surface, the scale of $L_0$ is in order of meter while the inner scale $l_0$ is typically in the order of a few millimeters. The number of random phase screens $M$ used to simulate atmospheric turbulence relies on both the distance of propagation and the intensity of the turbulence. For a given distance $\delta_z$, it depends on the Rytov variance whose mathematical expression is given as [23]

$$M \geq (10\sigma_R^2)^{6/11}, \tag{3}$$

Where $\sigma_R^2 = 1.23 C_n^2 k^{7/6} L^{11/6}$ is Rytov variance. The random phase screen is created by generating an $N \times N$ array of complex Gaussian random numbers $a + ib$, which are then multiplied by the square root of the phase spectrum. The phase spectrum ($\phi_\theta$) and power spectrum ($\phi_n$) are linked with following relation [24]

$$\phi_\theta(\kappa) = 2\pi k^2 \delta z \phi_n(\kappa), \tag{4}$$

where $\delta z$ is the spacing between the phase screens. Now we multiply the square root the phase spectrum with $\Delta_K^{-1}$, where $\Delta_K^{-1} = 2\pi/N\Delta$, $N$ is the number of sampling point and $\Delta$ is the spatial sampling interval, followed by the inverse Fourier transform. This method is known as the fast Fourier transform (FFT) method. The FFT method has a limitation in adequately sampling lower frequencies, due to which the generated phase screens do not include the phase contribution from spatial frequencies lower than the inverse of the screen width. Therefore, a significant portion of spectral energy near the origin remains almost unsampled. These lower frequencies play a central role in the beam centroid wandering. To resolve this, an additional sub-harmonic phase screen is incorporated to the FFT-generated random phase screen to include the missing lower frequencies [16,23]. The sub harmonic is generated by sampling the phase spectrum in neighborhood of $\kappa = 0$ using a significantly smaller sampling interval compared to the rest of the spectrum. The subharmonic screen method was first given by Lane et. al. [25] and later modified by Frehlich. [26]

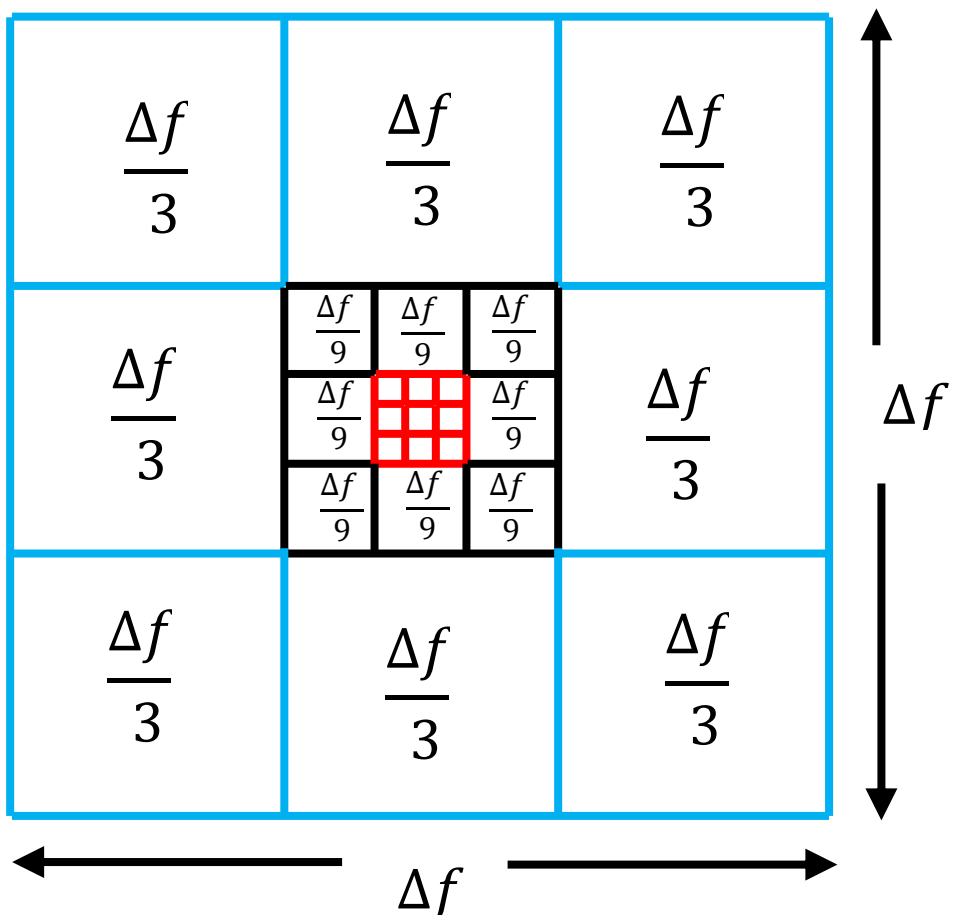

Fig. 1 subharmonic grid

The central pixel associated with spatial frequencies $f_x = 0$ and $f_y = 0$ in the high-frequency phase screen is subdivided into nine equally sized sub-pixels, as illustrated in the figure 1. This is the initial subharmonic level. Now, sampling points are placed within the eight surrounding sub-pixels, and the low-frequency screen is generated using the FFT method. The procedure is iteratively applied to the central small sub-pixel at the origin, generating multiple sub-harmonic levels as required. The size of the sample at the $p^{th}$ subharmonics are given by $\Delta f_{xp} = \Delta f_x / 3^p$. Now the phase contribution from each subharmonic is combined to get the total subharmonic phase screen.

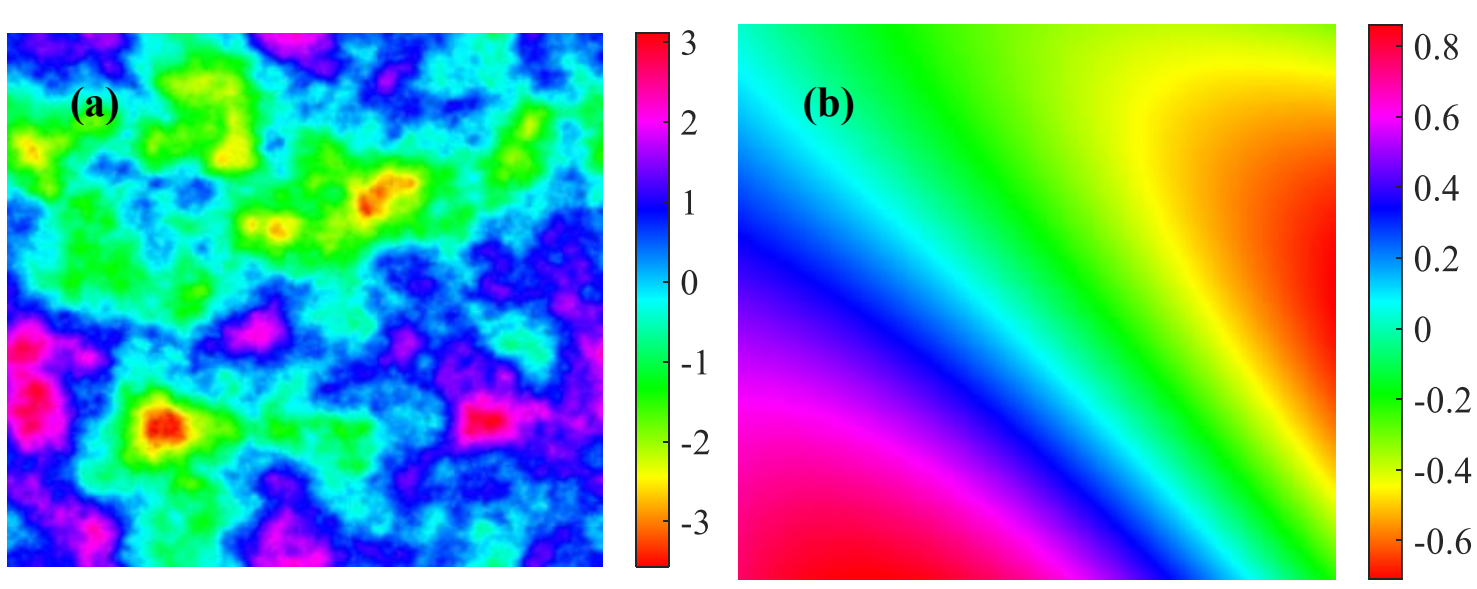

Fig. 2. Random phase screen using FFT method for $C_n^2 = 10^{-14} m^{-2/3}$ (a) random phase screen, and (b) subharmonic phase screen. Colorbars represent phase of the screens.

To unravel the dynamics of optical beams through atmospheric turbulence, we employ a sophisticated model featuring N random phase screens to approximate turbulence strength across the propagation distance $Z$. First the optical beam propagates a distance $\delta z$ using angular spectrum method followed by the multiplication with the phase transmittance $exp[i\theta(r)]$ of the random phase screen. This process is repeated until we reach the final distance $Z$.

### 3. Beam profiles and simulation parameters:

In this paper, we investigate the dynamics of AAF beams through atmospheric turbulence. Specifically, we explore two different types of AAF beams, first one is RAVB beam, and the other is RPVB beam. The mathematical expression of RAVB beam is given by [17]

$$E(r,\phi,0) = Ai\left(\frac{R_0-r}{w_0}\right)\exp\left(b\left(\frac{R_0-r}{w_0}\right)\right)\exp(il\phi), \tag{5}$$

where $Ai$ is the Airy function, $R_0$ is primary ring radius and $w_0$ is width of primary ring of the beam. The parameter $b$ is truncation factor here and $l$ is the topological charge [18]. While RPVB beam is expressed as [17],

$$E(r,\phi,0) = Pe\left(\frac{R_0-r}{w_0},\zeta_0\right) exp\left[b\left(\frac{R_0-r}{w_0}\right)\right] exp(il\phi), \tag{6}$$

where $Pe$ is the Pearcey function. The value of the parameter $\zeta_0$ is taken here as zero throughout the study. In this investigation, we have considered $w_0 = 3cm$, $R_0 = 7cm$ and $b = 0.1$. The outer scale $L_0 = 3m$ and the inner scale $l_0 = 1cm$ and the wavelength of the optical beam is $1550nm$.

### 4. Propagation and detection of OAM beam:

We now shift our focus to the multiplexing of different OAM modes and then their detection at the detector end. To achieve multiplexing of the different modes of AAF beam we simply add the complex mathematical expressions of these beams with different $l$ values. To detect the OAM modes at receiver, different types of methods have been proposed in the literature such as conjugate mode sorting [27], counting spiral fringes [28], measuring the Doppler effect [29], dove prism interferometers [30], and machine learning [31]. In this paper, we have used optical transformation sorting (log-polar method) for the detection of different modes [19,32].

#### 4.1 Optical transformation sorting:

Optical transformation sorting, developed by Lavery et. al. [33] refers to a technique employed in the field of optical signal processing for characterizing the various modes of the optical signal. This method enables the simultaneous detection of multiple (OAM) modes of the optical signals. In this method, we transform incoming OAM beam to log polar coordinate as

$$(x,y) \rightarrow (\rho,\theta) = \left(\log\left(\sqrt{x^2+y^2}\right), tan^{-1}\frac{y}{x}\right). \tag{7}$$

This method is motivated by the ability of a lens to focus a plane wave on its focal plane. The lateral position of this spot depends upon the transverse phase gradient of the plane wave. This enables the differentiation of multiple planes wave with different transverse phase gradients from one another. This transformation basically converts a ring into rectangle (see figure 3).

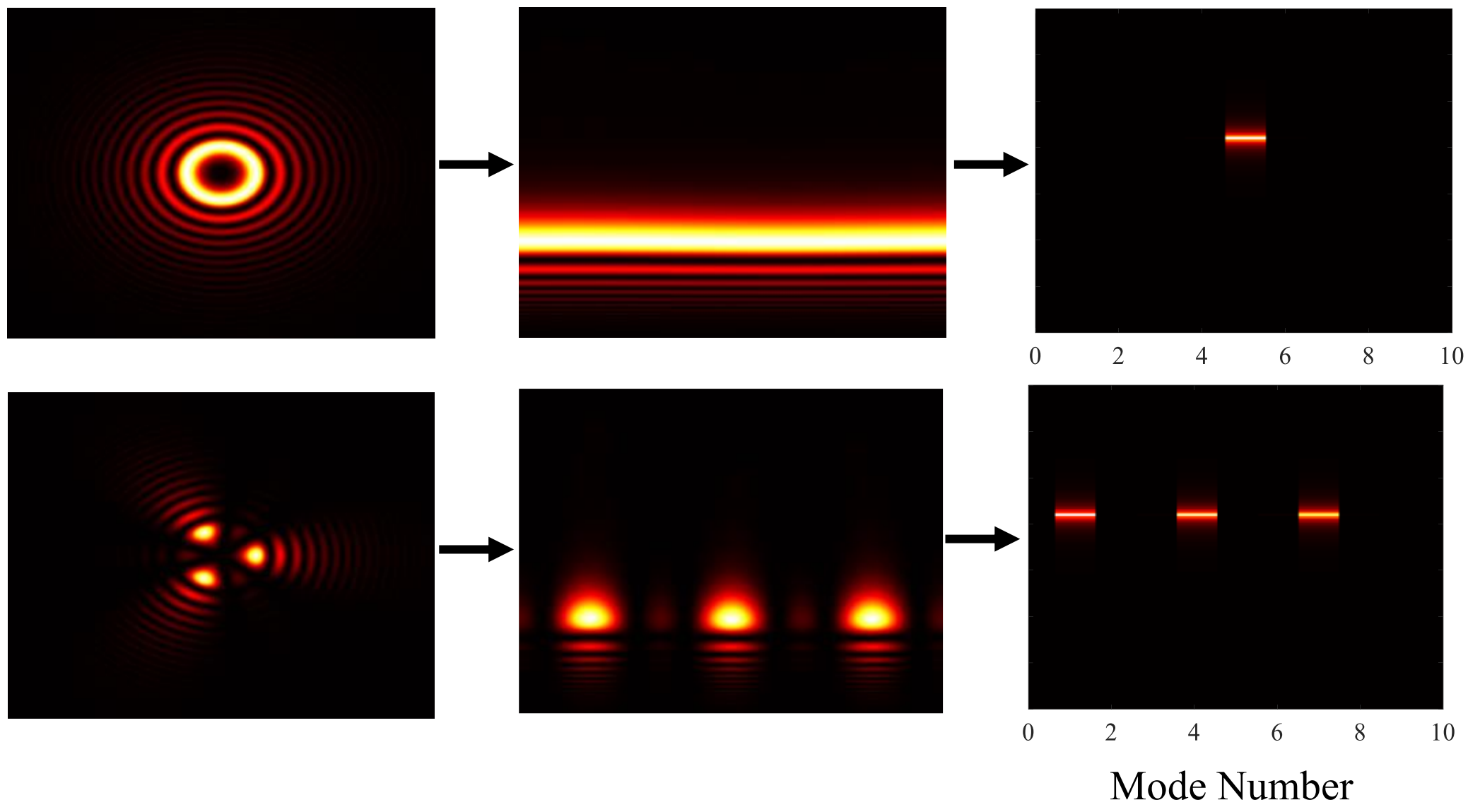

Fig. 3 The first row shows the mapping a RAVB of $l = 5$ in log polar coordinate. The second row shows the mapping of multiplexed RAVB of mode set [1,4,7] in log polar coordinate.

The first row in first column of figure 3 represents the transverse intensity profile of RAVB beam for $l = 5$ and the second row in first column of figure 3 represents the same for the mode set [1,4,7]. Similarly, the two rows in the second column of the figure shows the respective transverse profiles after transformation in log-polar coordinate. The third column of the figure shows the Fourier transform images of the second column data. The mapping transforms rotational and scaling effects into corresponding horizontal and vertical shifts. Following the above transformation, a Fourier transformation is applied in the new coordinate. By analyzing intensity of integer shift in the Fourier plane, one can detect the relative intensities across the various modes. In the context of non-multiplexed beams, determining the maximum Fourier intensity over the mode set gives the identification of the selected OAM mode. In the scenario of multiplexing, a critical threshold must be precisely selected to determine the presence or absence of a mode in the signal. In laboratory, such optical transformation system can be designed using two spatial light modulators with phase profiles [32]

$$\phi_1(x,y) = \frac{2\pi c}{\lambda f}\left[ytan^{-1}\left(\frac{y}{x}\right) - xlog\left(\frac{\sqrt{x^2+y^2}}{d}\right) + x\right], \tag{8}$$

and

$$\phi_2(x,y) = -\frac{2\pi cd}{\lambda f} exp\left(-\frac{u}{c}\right) exp\left(\frac{v}{d}\right), \tag{9}$$

where $c$ and $d$ are scaling and translating parameters. The phase profile $\phi_1$ transforms the beam into desired coordinate and $\phi_2$ is used for the phase correction. There are other methods for mode sorting, but we have chosen optical transformation sorting method because this method can simultaneously measure multiple states.

## 5. Propagation in atmospheric turbulence

Using random phase screen method mentioned in section 2, AAF beams have been made to propagate through multiple phase screens. We have considered two values for the turbulence strength namely $C_n^2 = 10^{-15}m^{-2/3}$ and $C_n^2 = 10^{-14}m^{-2/3}$ and the propagation distance $z = 1km$ has been used. As per the requirement of minimum number of phase screens, 20 random phase screens (satisfying eqn. (3)) have been used to accurately model the atmospheric turbulence. The screens were placed along the propagation distance at an interval of $\delta z = 100m$. The spatial window of size $1.2m \times 1.2m$ has been pixelated at $1024 \times 1024$. In this numerical study, we have performed 1000 iterations to provide sufficient statistics for calculation of channel efficiency and OAM spectrum.

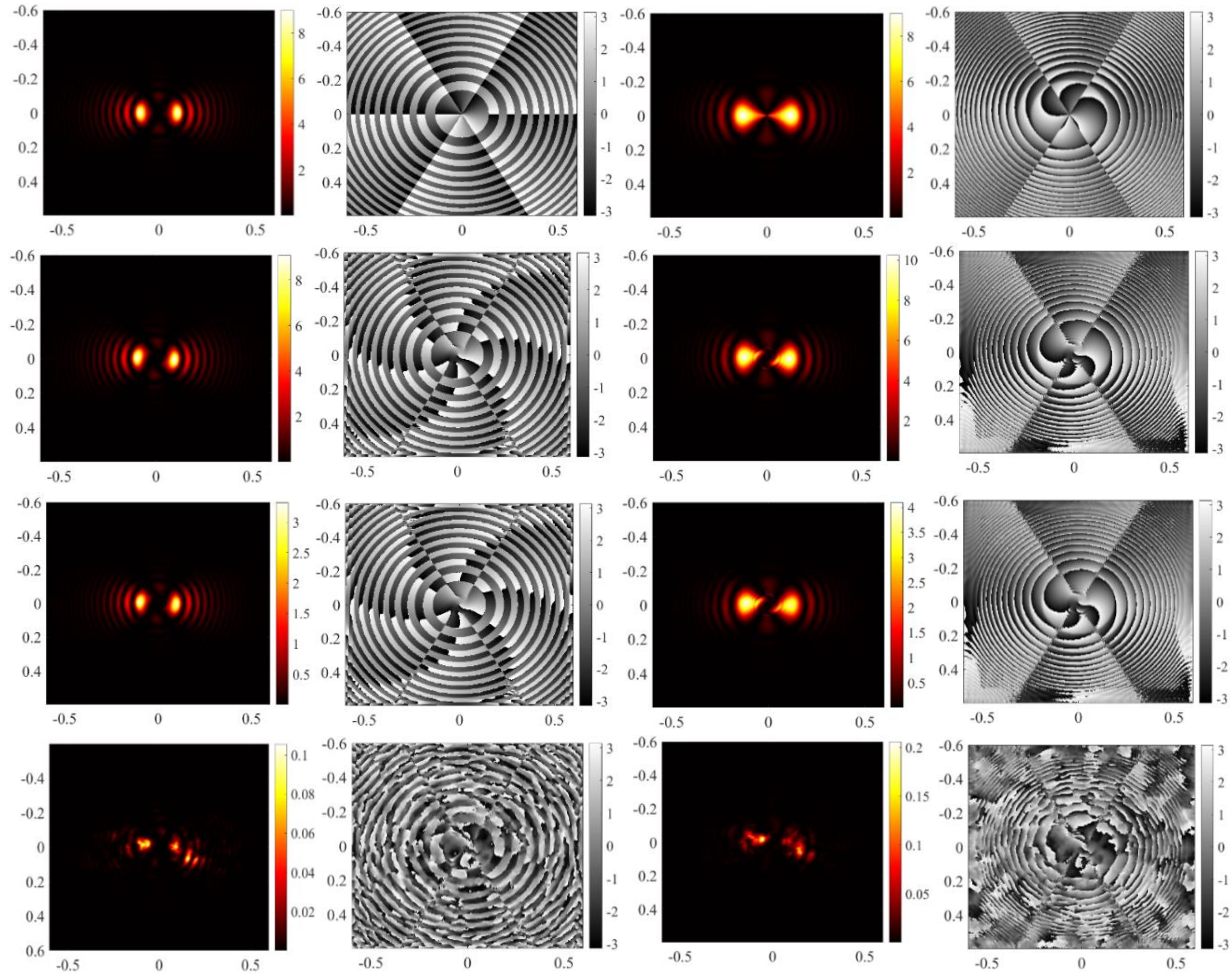

Fig. 4. Intensity and phase distributions of RAVB and RPVB in free space and turbulent atmosphere. The first row represents input multiplexed RAVB & RPVB of mode set [1, 3, 5]. Second row represents evolution in free space after $1km$ propagation distance. Third and fourth row represents evolution in atmospheric turbulence after propagating $1km$ for turbulence strength parameter $C_n^2 = 10^{-15}m^{-2/3}$ and $C_n^2 = 10^{-14}m^{-2/3}$ respectively. Horizontal and vertical axes are in meter (m).

We first investigate the effects of turbulence strength on multiplex RAVB and RPVB under similar conditions. We have implemented the angular spectrum method to calculate the beam profiles after propagation. We have chosen the mode set [1 3 5] and intensity proportion in these modes are kept identical. The first and the third columns in figure 4 show the transverse intensity profiles of RAVB and RPVB respectively. While the second and the fourth columns show the transverse phase profiles of RAVB and RPVB respectively. These intensity and phase profiles are plotted after $1km$ distance of propagation except for first row of the figure. The first row in the figure 4 represents input multiplexed RAVB & RPVB of mode set [1, 3, 5]. Second row represents evolution in free space after $1km$ propagation distance. Third and fourth row represents evolution in atmospheric turbulence after propagating $1km$ for turbulence strength parameter $C_n^2 = 10^{-15}m^{-2/3}$ and $C_n^2 = 10^{-14}m^{-2/3}$ respectively. From the second row it can been that these beams maintain their transverse intensity profile over the specified distance due to the self-healing and non-diffracting nature of the beams. As can be seen in figure 4, for stronger turbulence ($C_n^2 = 10^{-14}m^{-2/3}$) intensity and phase profiles of the beams show a transition towards a chaotic pattern that ultimately leads to the loss of the self-healing and non-diffracting characteristics of the beams. Hence, here onwards we only consider turbulence strength weaker than $10^{-14}m^{-2/3}$. Turbulence strengths smaller than $C_n^2 = 10^{-14}m^{-2/3}$ are usually defined as weak turbulence.

## 6. Channel efficiency:

Channel efficiency can be defined by analyzing the energy detected in each demultiplexed mode. To calculate the channel efficiency, we calculate the energy contained in the demultiplexed mode via optical transformation sorting method. Channel efficiency of non-multiplexed mode have been calculated as

$$\text{Channel efficiency non-multiplexed beam} = \frac{energy\ of\ the\ detected\ mode\ at\ the\ output}{Input\ \ beam\ energy} \quad (10)$$

Figure 5 illustrates the impact of different levels of atmospheric turbulence after propagating $1km$ for $C_n^2 = 10^{-16}m^{-2/3}$ and $C_n^2 = 10^{-15}m^{-2/3}$ on channel efficiency for RAVB and RPVB with different OAM mode number. In figure 5(a), we see that upon increasing the value of topological charge *l,* the channel efficiency effectively decreases for both the beams. This happens, because larger value of *l* increases the transverse profile of the optical beam leading to substantial interactions with atmospheric turbulence ($C_n^2 = 10^{-16}m^{-2/3}$), and most of the energy will be transferred to the other modes, which ultimately leads to reduction in the channel efficiency. Similar kind of observations are there in figure 5(b) where $C_n^2 = 10^{-15}m^{-2/3}$. We also notice that upon increasing the turbulence strength channel efficiency for both beam decreases as per the aforementioned explanation. These figures also reveal that as the value of *l* increases, the channel efficiency of RAVB surpasses that of RPVB across all the mentioned atmospheric strengths and mode numbers. This suggests that opting RAVB over RPVB for FSO is advisable, as it maximizes the received energy of each mode at the receiver end.

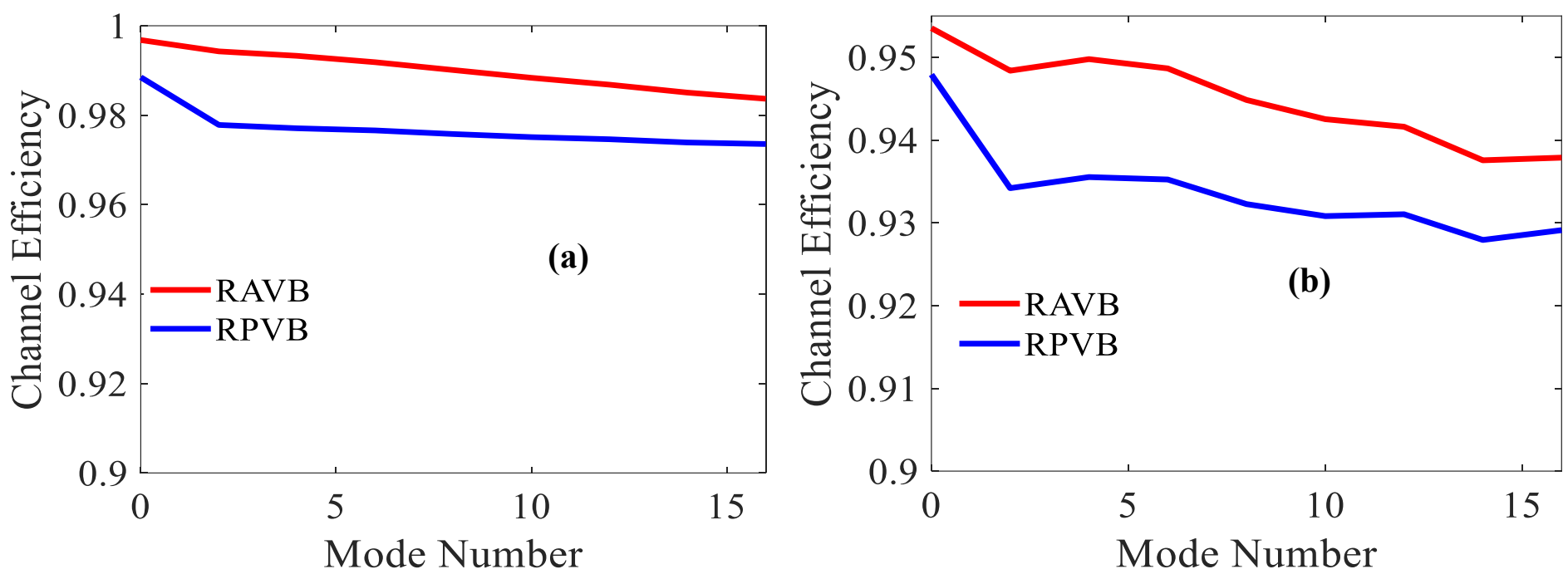

Fig. 5. Channel efficiency of RAVB and RPVB in atmospheric turbulence after $1km$ of distance of propagation. In (a) $C_n^2 = 10^{-16}m^{-2/3}$ and (b) $C_n^2 = 10^{-15}m^{-2/3}$.

### 6.1 Channel efficiency of multiplexed beam

To evaluate the channel efficiency of multiplexed RAVB and RPVB beams in atmospheric turbulence, we first combined these beams and then resulting multiplexed beam in varying levels of atmospheric turbulence is propagated over a distance of *1 km*. At the receiver end we calculate the energy of each demultiplexed mode in the mode set under consideration. We have defined here the formula for the calculation of channel efficiency of multiplexed beam as ratio of sum of individual energy of demultiplexed mode at output to total energy the input multiplexed beam-

$$\text{Channel efficiency of multiplexed beam} = \frac{Sum\ of\ indivdual\ energy\ of\ demultiplexed\ mode\ at\ the\ output}{Input\ mutiplexed\ beam\ energy} \quad (11)$$

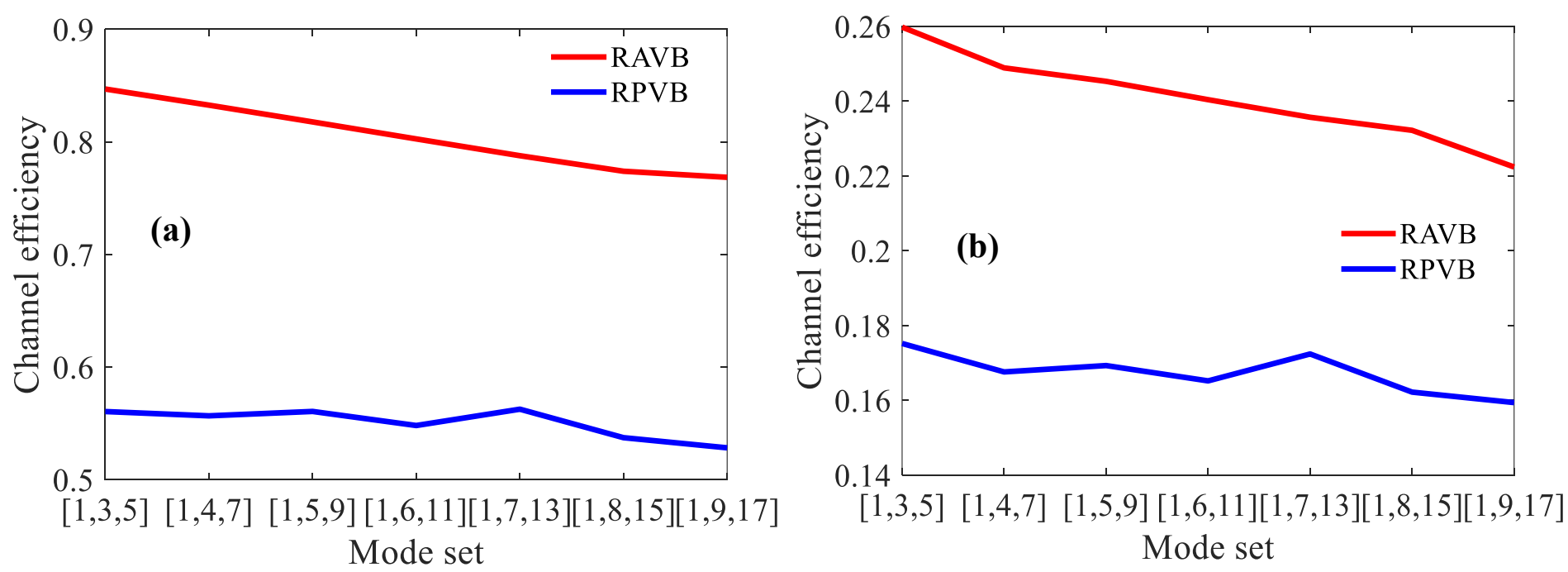

Fig. 6. Channel efficiency of multiplexed RAVB and RPVB in atmospheric turbulence after $1km$ of distance of propagation. (a) $C_n^2 = 10^{-16}m^{-2/3}$ and (b) $C_n^2 = 10^{-15}m^{-2/3}$.

We have considered here seven sets of three OAM modes as [1, 3, 5], [1, 4, 7], [1, 5, 9], [1, 6, 11], [1, 7, 13], [1, 8, 15], [1, 9, 17] with increasing mode spacing for the multiplexing of RAVB and RPVB. We see from figure 6 that RAVB outperforms RPVB a result similar to single mode efficiency investigation. It can also be seen from the same figure that with increase in the separation between the OAM modes, channel efficiencies for both the beams decrease effectively. This happens because the turbulent atmosphere causes the OAM mode to spread the energy to the adjacent modes.

## 7. OAM spectrum

In this section, we numerically study the OAM spectra of the AAF beams under consideration in atmospheric turbulence. To identify OAM modes at receiver end, we have used optical transformation sorting (log-polar) method as discussed in the previous section. Figure (7) shows the OAM spectra of RAVB and RPVB with mode $l = 1$ after propagating $3\ km$ in different levels of atmospheric turbulence. The first row in figure (7) depicts OAM spectra of RAVB and second row of RPVB. In first column strength of the turbulence $C_n^2 = 10^{-16}m^{-2/3}$, in second column $C_n^2 = 10^{-15}m^{-2/3}$ and in third column $C_n^2 = 5 \times 10^{-15}m^{-2/3}$. We see from figure 7(a) and figure 7(d) that there is no spreading of the topological charge $l = 1$ for $C_n^2 = 10^{-16}m^{-2/3}$ for both the beams. However, upon increasing the turbulence strength, we observe crosstalk among the OAM modes (see figures 7(b, c, e & f)). Crosstalk is the energy detected by receiver for which transmitted signal is not intended. In figure 7(b) and (e)) we see some crosstalk among the adjacent mode for $C_n^2 = 10^{-15}m^{-2/3}$. In intense turbulence level ($C_n^2 = 5 \times 10^{-15}m^{-2/3}$), we observe more crosstalk among other modes and more energy gets dispersed among other modes for both beams (see figures 7(c) and 7(f)). We also observe that, in turbulent atmosphere, the energy associated with the transmitted OAM mode decreases as turbulence strength increases. This reduction in energy becomes more pronounced with stronger turbulence.

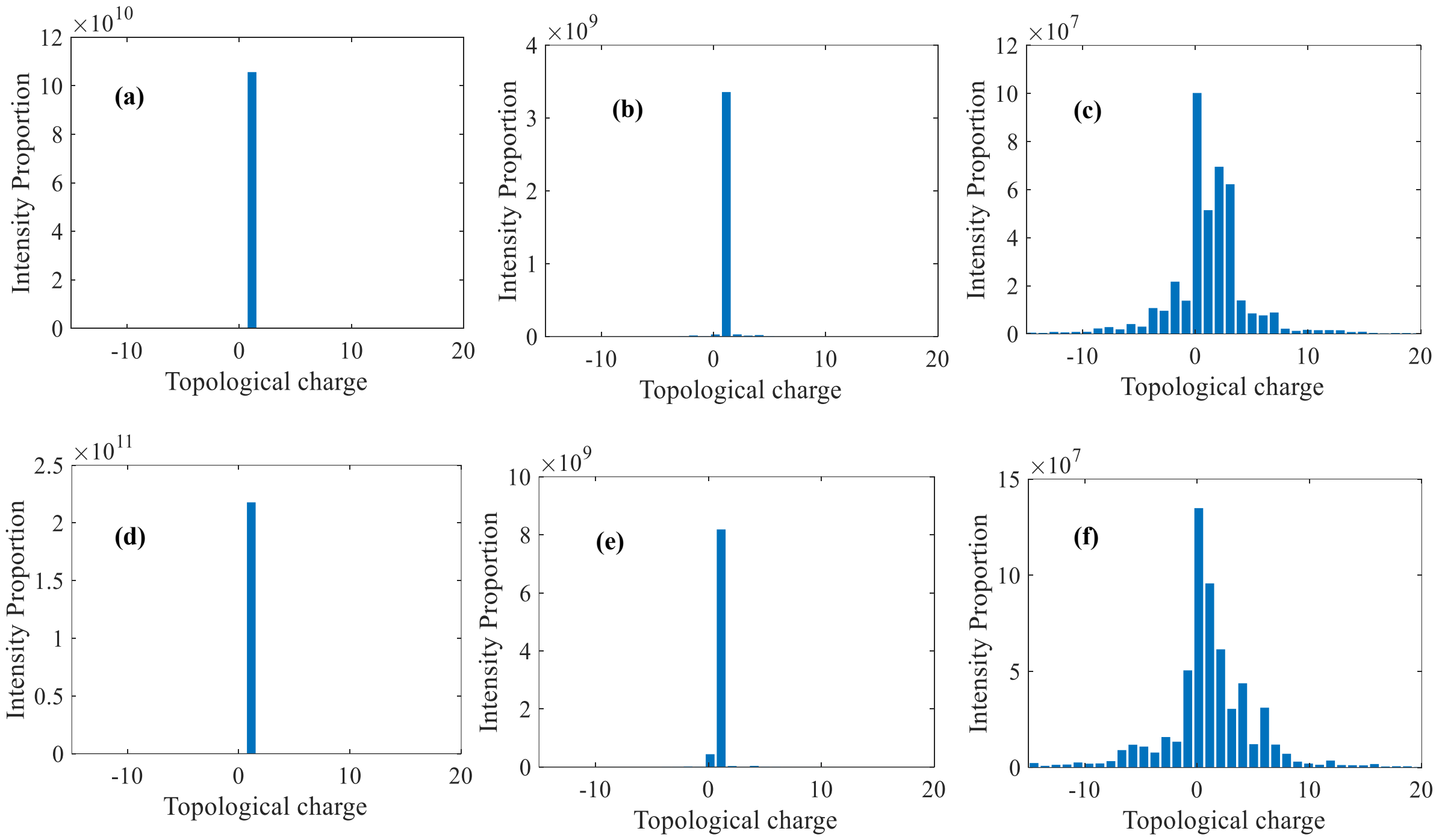

Fig. 7. Distribution of OAM spectrum of RAVB and RPVB in different levels of atmospheric turbulence. The first row represents OAM spectrum of RAVB and second row of RPVB after propagating $3km$ through the atmospheric turbulence. In first column $C_n^2 = 10^{-16}m^{-2/3}$, in second column $C_n^2 = 10^{-15}m^{-2/3}$ and in third column $C_n^2 = 5 \times 10^{-15}m^{-2/3}$.

In figure (8), we have studied the evolution of OAM spectra of RAVB and RPVB at different propagation distances for $C_n^2 = 5 \times 10^{-15}m^{-2/3}$. In the figure, the first row represents OAM spectrum of RAVB while second row is for RPVB beam. First column of the figure represents spectra at $1km$, second column at $2km$ while the third column is at $4km$. In figure 8(a) and 8(d), we see that crosstalk is negligible at $1km$ propagation distance for both the beams. However, when we move further in turbulent atmosphere as depicted in figure 8(b) and figure8(e) crosstalk emerges with energy dispersing to other adjacent modes. Moving further in turbulent atmosphere (figure 8(c) and 8(f)), we observe more energy dispersion to the adjacent modes i.e. crosstalk increases on increasing the propagation distance in turbulent atmosphere. This happens because the effect of the turbulence becomes stronger with larger propagation distance. For quantitative comparison of OAM spectra of RAVB and RPVB, we have calculated the weighted standard deviations for figures 7(c) and 7(f). The weighted standard deviations for the two figures are calculated respectively to be 3.47, and 3.56. Weighted standard deviations for figures 8(c) and 8(f) are 4.21, and 4.34 respectively. This confirms that RPVB is more dispersive as compared to RAVB.

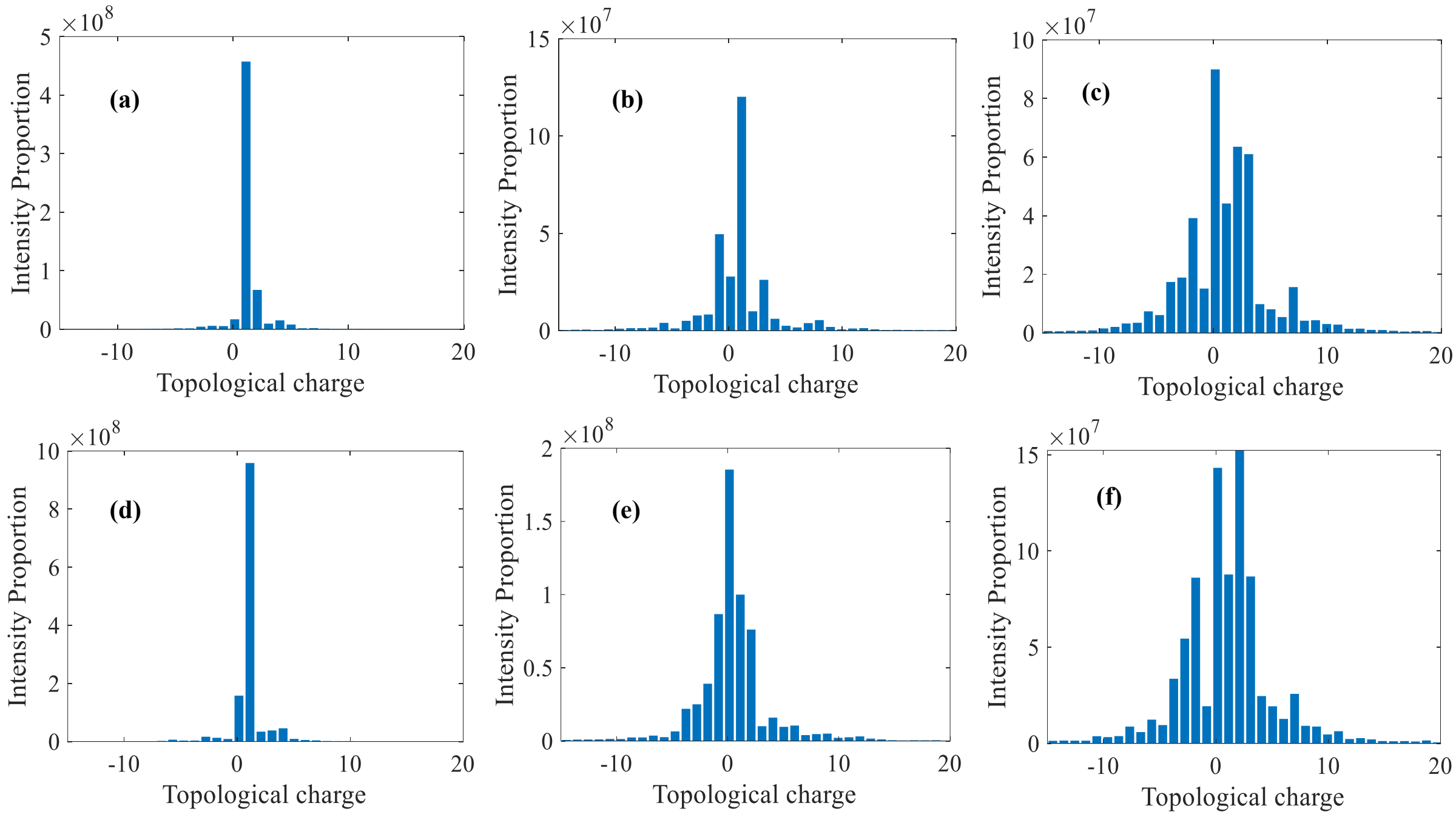

Fig. 8. Distribution of OAM spectrum of RAVB and RPVB at different propagation distance in $C_n^2 = 5 \times 10^{-15} m^{-2/3}$. The first row represents the OAM spectrum of RAVB and second row of RPVB beam. OAM spectra as shown in first, second and third columns are plotted respectively after propagating $1km$, $2km$ and $4km$ in turbulent atmosphere.

## 8. Conclusion

In summary, we have studied the evolution of multiplexed RAVB and RPVB in turbulent atmosphere and compared their channel efficiency and OAM spectra by observing their transverse intensity and phase profiles' dynamics. In the case of single mode channel, under weak atmospheric turbulence we have observed a decrease in channel efficiency for both the beams as the mode number was increased. Furthermore, the study revealed that for multiplexed RAVB and RPVB, channel efficiency decreases with increase in spacing between mode numbers in the mode sets. The analysis highlighted the superior performance of RAVB over RPVB in weak atmospheric turbulence. We have further studied OAM spectra of RAVB and RPVB and reported an increase in crosstalk among different OAM modes as turbulence strength was increased for both the beams. Moreover, we have observed a consistent increase in crosstalk among OAM modes as the propagation distance increases in a given turbulent atmospheric condition. On comparing the OAM dispersion for these two beams, we have concluded that the topological charge associated with RPVB disperses relatively strongly as compared to that for RAVB.

We believe that our findings may significantly contribute towards the understanding of the structured light based FSO system. The results of the study may hold applications in satellite communication and free space sensing.

## 9. Acknowledgement

Shakti Singh wishes to acknowledge the University Grant Commission (India) for the financial support. This work is supported by the research grant CRG/2022/007736 from SERB (India). The authors are also thankful to Prof. Anshul Jaiswal from the department of electronics and communication engineering IIT Roorkee for the fruitful discussion on the calculation of channel efficiency of the multiplexed optical beams.

**Disclosure.** The authors declare no conflicts of interest.

**Data availability.** Data underlying the results presented in this paper are not publicly available at this time but may be obtained from the authors upon reasonable request.